\documentclass[12pt]{article}
\usepackage{array}
\usepackage{graphicx}
\usepackage{amssymb}
\usepackage{amsmath}
\input{epsf}
\usepackage{cite}
\def\@fmsl@sh#1#2#3{\m@th\ooalign{$\hfil#1\mkern#2/\hfil$\crcr$#1#3$}}
 \def\eq#1\en{\begin{equation}#1\end{equation}}
\def\s[#1,#2]{[#1\stackrel{\star}{,}#2]}
\def\sx[#1,#2]{[#1\stackrel{\star_{x}}{,}#2]}

\newcommand{\nc}{\newcommand}
\nc{\beq}{\begin{equation}}
\nc{\eeq}{\end{equation}}
\nc{\beqa}{\begin{eqnarray}}
\nc{\eeqa}{\end{eqnarray}}

\def\bc{\begin{center}}
\def\ec{\end{center}}

\def\to{\rightarrow}

\def\gsim{\mathrel{\mathpalette\atversim>}}

\def\bc{\begin{center}}
\def\ec{\end{center}}

\def\gsim{\mathrel{\rlap{\lower4pt\hbox{\hskip1pt$\sim$}}

    \raise1pt\hbox{$>$}}}       %greater than or approx. symbol

\def\gsim{\mathrel{\rlap{\lower4pt\hbox{\hskip1pt$\sim$}}
    \raise1pt\hbox{$>$}}}       %greater than or approx. symbol

\begin{document}
\makeatletter
\def\fmslash{\@ifnextchar[{\fmsl@sh}{\fmsl@sh[0mu]}}
\def\fmsl@sh[#1]#2{%
  \mathchoice
    {\@fmsl@sh\displaystyle{#1}{#2}}%
    {\@fmsl@sh\textstyle{#1}{#2}}%
    {\@fmsl@sh\scriptstyle{#1}{#2}}%
    {\@fmsl@sh\scriptscriptstyle{#1}{#2}}}
\def\@fmsl@sh#1#2#3{\m@th\ooalign{$\hfil#1\mkern#2/\hfil$\crcr$#1#3$}}
\makeatother
%\baselineskip 24pt

%%%%%%%%%%%%%%%%%%%%%%%%%%%%%%%%%%%%%%%%%%%%%%%%%%%%%%%%%%%%%%%%%
%%%
%%%                      TITLE PAGE
%%%
%%%%%%%%%%%%%%%%%%%%%%%%%%%%%%%%%%%%%%%%%%%%%%%%%%%%%%%%%%%%%%%%%
\thispagestyle{empty}
\begin{titlepage}
\boldmath
\begin{center}
  \Large {\bf Remarks on Higgs Inflation}
  \end{center}
\unboldmath
\vspace{0.2cm}
\begin{center}
{{\large Michael Atkins}\footnote{m.atkins@sussex.ac.uk} and 
{\large Xavier Calmet}\footnote{x.calmet@sussex.ac.uk}
}
 \end{center}
\begin{center}
{\sl Physics and Astronomy, 
University of Sussex,  \\ Falmer, Brighton, BN1 9QH, UK 
}
\end{center}
\vspace{\fill}
\begin{abstract}
\noindent
We discuss models where the Higgs boson of the electroweak standard model plays the role of the inflaton. We focus on the question of the violation of perturbative unitarity due to the coupling of the Higgs boson either to the Ricci scalar or to the Einstein tensor and discuss the background dependence of the unitarity bounds.  Our conclusion is that the simplest model which restricts itself to the standard model Higgs boson without introducing further degrees of freedom has a serious problem. However, in the asymptotically safe gravity scenario, the Higgs boson of the standard model could be the inflaton and no physics beyond the standard model is required to explain both inflation and the spontaneous breaking of the electroweak symmetry of the standard model.

\end{abstract}  
\end{titlepage}

%\pacs{}

%%%%%%%%%%%%%%%%%%%%%%%%%%%%%%%%%%%%%%%%%%%%%%%%%%%%%%%%%%%%%%%%
%%%
%%%                    MAIN TEXT
%%%
%%%%%%%%%%%%%%%%%%%%%%%%%%%%%%%%%%%%%%%%%%%%%%%%%%%%%%%%%%%%%%%%

\newpage

\section{Introduction}

It is now widely accepted that the early universe went through a period of rapid inflation. It has recently been proposed that the standard model Higgs boson could be the inflaton \cite{CervantesCota:1995tz, Bezrukov:2007ep, DeSimone:2008ei}. The motivation is clear. The standard model needs a fundamental scalar field to break the electroweak symmetry and inflation is well described by scalar fields. Why not identify both scalar fields? While being an elegant and minimal model, there has been much debate over whether this scenario suffers from unitarity problems during the inflationary period \cite{Lerner:2009na, Bezrukov:2010jz, Burgess:2009ea, Burgess:2010zq, Barbon:2009ya, Atkins:2010eq, Atkins:2010re, Han:2004wt, Hertzberg:2010dc, Ferrara:2010in}. It has recently been emphasized that the scale of unitarity violation depends on the size of the background fields in the model \cite{Ferrara:2010in, Bezrukov:2010jz}. Since the Higgs field is required to have a large background value during inflation the unitarity bound is higher during the inflationary period and the theory does not violate unitarity at any time. Despite this claim, we point out in this paper that the requirement of new physics appearing below the Planck mass in today's universe sacrifices the original idea of the standard model alone being able to explain inflation. In fact despite the theory being unitary at all times it is most likely that new physics which is required to solve the unitarity problem in today's universe will take part in the inflationary dynamics and it will no longer be driven by the Higgs boson alone.

In an attempt to bypass the unitarity problem, a new model of Higgs inflation has been introduced with a derivative coupling to gravity \cite{Germani:2010gm}. In the second part of this paper we carry out a detailed analysis of the scale of unitarity violation in this model and find that contrary to the initial claims, it also suffers from unitarity problems.  

Previous attempts to unitarize Higgs inflation have required the introduction of new degrees of freedom \cite{Lerner:2009xg, Giudice:2010ka} or higher dimensional operators \cite{Lerner:2010mq}. In the last part of this paper we note that if gravity were to exhibit asymptotic safety in the presence of the couplings which drive Higgs inflation then the models might not suffer from any unitarity issues. This would provide a fully consistent paradigm for Higgs inflation without having to introduce any new degrees of freedom beyond those found in the standard model.

\section{Original Model of Higgs Inflation}

The original model of Higgs inflation proposed in \cite{Bezrukov:2007ep} is given by the action 
\begin{equation}
\label{action1}
S=\int d^4x \sqrt{-g} \left[ \frac{\bar M_P^2}{2}R - \xi {\cal H^\dagger H } R + {\cal L}_{SM} \right] ,
\end{equation}
where $\cal H$ is the standard model Higgs doublet, ${\cal L}_{SM}$ is the standard model Lagrangian and $\bar M_P $ is the reduced Planck mass. $\xi$ is a new coupling constant which needs to be of the order $10^4$ to produce successful inflation.

Such a large coupling has raised doubts about the validity of the theory. It has been well established \cite{Burgess:2009ea, Barbon:2009ya, Atkins:2010eq, Atkins:2010re, Han:2004wt, Burgess:2010zq, Hertzberg:2010dc, Ferrara:2010in} that when the  Higgs field is expanded around the standard model vacuum expectation value, $v=246$ GeV, the theory violates unitarity at a scale $\Lambda \simeq \bar M_P / \xi$. This scale should be considered as a cut off for the effective theory. The size of the Higgs field during inflation is above $\bar M_P/\sqrt{\xi}$ which is above the cut off, putting the inflationary calculations in jeopardy.

In \cite{Ferrara:2010in, Bezrukov:2010jz} it is shown that the cut off is in fact dependent on the size of the background fields. Since the Higgs field has a large background value during inflation it is found that throughout the period of inflation and reheating the fields lie below this background dependent cut off. While this theoretically may allow one to work within the regime of validity of the effective theory for inflationary calculations, we wish to point out that the original idea of the standard model alone being able to explain inflation has to be sacrificed.

The original idea of Higgs inflation was so exciting because it did not require the introduction of new physics below the Planck mass to explain particle physics and inflation. We do of course expect new physics to appear at the gravitational cutoff, $\sqrt{\bar M_P^2 + \xi \bar\phi^2}$ where $\bar\phi$ is the background Higgs field \cite{Bezrukov:2010jz}. However, if we require that the standard model alone be valid up to the gravitational cutoff today, i.e. $\bar M_P$, we get a bound on the size of $\xi$ ($-0.81 \le \xi \le 0.64$, see \cite{Atkins:2010eq, Atkins:2010re}). Since $\xi$ is time independent, this bound restricts $\xi$ for all time and we again see that the original idea of Higgs inflation is inconsistent. 

Given that the action (\ref{action1}) represents a non-renormalizable effective theory, we take the point of view that it is only valid in an energy regime up to a certain cut off, $\Lambda$. New physics, characterized by a mass scale above this cut off, has been integrated out and appears as higher dimensional operators suppressed by powers of $\Lambda$. In the case of Higgs inflation these higher dimensional operators will take the form  
\begin{equation}
\label{operators}
 \frac{({\cal H^\dagger H})^n R^m}{\Lambda^{2(n+m)-4}}  .
 \end{equation}
Since there is a violation of unitarity at the scale $\Lambda \simeq \bar M_P / \xi$ when the fields are expanded around small field values, we know that in today's universe, new physics must appear, characterized by a mass scale $m \sim \bar M_P / \xi$, in order to fix the unitarity problem. The original minimalistic view of the standard model Higgs boson alone, valid up to $\bar M_P$, explaining inflation has to be given up. Further, since the scale of unitarity violation increases with large background field values, either this new physics remains at this mass scale during inflation or any mechanism that lifts the mass scale of the new physics would likely mix the new degrees of freedom with the Higgs boson, as happens in \cite{Giudice:2010ka}. Either way, the new physics will become a part of the inflationary dynamics, not just degrees of freedom required to unitarize the model. Indeed, in \cite{Giudice:2010ka} it is the presence of the large non-minimal coupling to the new sigma field that is of critical importance for inflation, not the presence of the Higgs boson.

We would like to make a brief comment here on the use of singlet scalars non-minimally coupled to gravity. As has now been well established, the amplitude for $2 \to 2$ gravitational scattering of a singlet scalar must necessarily include $s$, $t$ and $u$-channel processes. When this is done, a cancellation amongst these diagrams occurs, raising the scale of unitarity violation to $\Lambda \sim \bar M_P$ \cite{Huggins:1987ea}. However as was pointed out in \cite{Hertzberg:2010dc} and in a footnote in \cite{Barbon:2009ya} the presence of a self coupling term (e.g. $\lambda \phi^4$) in the potential for the singlet field will introduce a violation of unitarity at a scale $\Lambda \sim \bar M_P / (\sqrt{\lambda}\xi)$. This comes from the $(\lambda \xi / \bar M_P ) \phi^6$ term in the Einstein frame action. Since there has been some confusion in the past over the scale of unitarity violation in the Jordan and Einstein frames we would like to point out here where this problem also appears in the Jordan frame. In the Jordan frame we consider $2 \to 4$ scattering at order ${\cal O} (\lambda \xi^2 / \bar M_P^2)$ and find a cross section $\sigma \sim \lambda^2 \xi^4 s / \bar M_P^4$, where $s$ is is the center of mass energy squared. Unitarity requires $\sigma < 8\pi /s$, giving a scale of unitartiy violation $\Lambda \sim \bar M_P / (\sqrt{\lambda}\xi)$. In \cite{Hertzberg:2010dc} it is suggested that if $2 \to n$ scattering is considered in the large $n$ limit, $\Lambda \to \bar M_P / \xi$, which is independent of $\lambda$. However this fails to take into account the full phase space integration, which, as pointed out in \cite{Dicus:2004rg}, includes factors of $1/ (n-1)!(n-2)!$ that dominate in the large $n$ limit. The exact process which generates the lowest bound (i.e. $2 \to 4$, $2 \to 6$ or $2\to 8$ etc.) is dependent on the size of $\lambda$ but the scale of unitarity violation does not vary much in these cases. Since we are anyway only presenting a power counting argument, we cannot accurately determine the exact point of unitarity breakdown which would require the calculation of the full amplitudes. Hence we maintain that for $\lambda<1$ the scale of unitarity violation is still $\Lambda \sim \bar M_P / (\sqrt{\lambda}\xi)$ and the size of $\lambda$ is significant as one would expect.  Models which use a singlet scalar non-minimally coupled to gravity therefore need to carefully choose their self couplings, and indeed couplings to other sectors of the model, or require further new physics to unitarize the model.

New physics beyond the standard model taking part in the inflationary dynamics seems to go against the original idea of using the Higgs boson both for inflation and to break the electroweak symmetry of the standard model by using only one fundamental scalar field. In section \ref{asym} of this paper we discuss the possibility of the unitarity problem being completely negated if gravity is asymptotically safe and without having to introduce any new physics beyond the standard model. Let us first, however, reconsider an interesting idea to avoid the unitarity problem by using derivative couplings of the Higgs boson to the Einstein tensor instead of the ${\cal H^\dagger H } R$ coupling.

\section{New Model of Higgs Inflation}

To overcome the unitarity problems associated with the original proposal for Higgs inflation, Germani and Kerhagis proposed a new model where the Higgs boson has a derivative coupling to the Einstein tensor \cite{Germani:2010gm}. They claimed that this new model was free of unitarity problems and could produce  successful inflation. In a later paper \cite{Germani:2010ux} they calculated the cosmological perturbations in the model and showed that they were consistent with the latest WMAP data. Since the prime motivation for the new model was to overcome the unitarity problems associated with the earlier model, it is important to carry out a thorough analysis of the scale of unitarity violation in this model. We find that contrary to the original claims, the new model of Higgs inflation also suffers from unitarity problems during the inflationary period.

In \cite{Germani:2010gm} it is shown that the unique non-minimal derivative coupling of the Higgs boson to gravity, propagating no more degrees of freedom than general relativity minimally coupled to a scalar field, is given by the action
\begin{equation}
\label{action}
S=\int d^4x \sqrt{-g} \left[ \frac{\bar M_P^2}{2}R - \frac12 (g^{\mu\nu}-w^2 G^{\mu\nu})\partial_\mu \Phi \partial_\nu \Phi - \frac{\lambda}{4} \Phi^4 \right] ,
\end{equation}
where $G^{\mu\nu} = R^{\mu\nu}-\frac{R}{2} g^{\mu\nu}$ is the Einstein tensor, $w$ is an inverse mass scale, and $\Phi$ represents a real scalar field which is to be associated with one of the real degrees of freedom of the standard model Higgs doublet.

To calculate the scale at which unitarity is violated in such a theory we consider $\Phi \Phi \to \Phi \Phi$ scattering via graviton exchange. As in \cite{Atkins:2010eq, Han:2004wt}, we simplify the calculation by only considering s-channel scattering, we can do this by imposing different in and out states, i.e. $\Phi \Phi \to \Phi^\prime \Phi^\prime$. This is justified for the case of the standard model Higgs doublet, which in the high energy regime being considered, appears as four real scalars. Expanding around the inflating background $g_{\mu\nu}=\bar g_{\mu\nu} + h_{\mu\nu}/\bar  M_P$ where $\bar g_{\mu\nu}={\rm diag}(-1,a(t),a(t),a(t))$ is the Friedmann-Roberstson-Walker (FRW) metric, to lowest order in $h_{\mu\nu}$ the Einstein tensor is $G_{\mu\nu}=-3H^2 \bar g_{\mu\nu}$ where $H\equiv\dot{a}/a$ is the Hubble constant. For $wH\gg1$, expanding $\Phi$ around its background value during inflation, $\Phi_0$, we have $\Phi=\Phi_0 + \frac{1}{\sqrt{3}wH}\phi$ where $\phi$ is canonically normalized. As in  \cite{Germani:2010gm}, we find an interaction term
\begin{equation}
\label{interaction}
I \simeq \frac{1}{2 H^2 \bar M_P} \partial^2 h^{\mu\nu}\partial_\mu \phi \partial_\nu \phi. 
\end{equation} 
A power counting analysis then gives the scale at which unitarity is violated to be
\begin{equation}
\label{interaction2}
\Lambda \simeq (2  H^2 \bar M_P )^{1/3}. 
\end{equation}
In \cite{Germani:2010ux}, by direct comparison with the WMAP data and considering the allowed range of the standard model Higgs boson self coupling, the size of the background fields during inflation are found to be
\begin{equation}\label{Rvalue}R \simeq 5.6 \times 10^{-8} \bar M_P^2 \; ,\end{equation}
\begin{equation}\label{Phivalue} 2.1 \times 10^{-2} \bar M_P < \Phi_0 < 2.7 \times 10^{-2} \bar M_P \;. \end{equation}
In order for the higher dimensional operators (\ref{operators}) to be suppressed we must ensure that during inflation, $R<\Lambda^2$ and $\Phi_0 < \Lambda$. We can determine $H\simeq \sqrt{R/12}$ from (\ref{Rvalue}) and we find
\begin{equation}\label{lambdavalue}\Lambda \simeq 2 \times 10^{-3} \bar M_P.\end{equation}
In \cite{Germani:2010gm} only the condition $R<\Lambda^2$ was considered and the model was said to be free of unitarity problems. However, considering the Higgs field (\ref{Phivalue}) we see that $\Phi_0 > \Lambda$ during inflation and the model in fact suffers from unitarity problems.

It is also of interest to calculate $\Lambda$ around today's background since this gives us the lowest energy at which new physics must appear in order to unitarize the theory. Expanding around a flat background $g_{\mu\nu}=\eta_{\mu\nu}+\sqrt{2}h_{\mu\nu}/\bar M_P + {\cal O}(\bar M_P^{-2})$ and the standard model Higgs boson vacuum expectation value (which we take to be zero in the high energy limit being considered), the cut off is found to be
\begin{equation}\label{lambda2}\Lambda \simeq   \left( \frac{5 \bar M_P}{w^2} \right)^{1/3}.\end{equation}
In \cite{Germani:2010ux}, by comparison with the WMAP data the value of the dimensionful parameter $w$ is found to lie in the range
\begin{equation}\label{wvalue}7 \times 10^{-8} \: \bar M_P < w^{-1} < 8.8 \times 10^{-8} \: \bar M_P . \end{equation}
Taking the upper bound for $w^{-1}$ we find that unitarity is violated at
\begin{equation}\label{bound}\Lambda = 3.4\times10^{-5}\bar M_P \end{equation}
which is smaller than both $\sqrt{R}$ and $\Phi_0$ during inflation.

We conclude that, during the inflationary period, new physics must be present to cure the unitarity problem and would likely spoil the inflationary potential.

\section{Asympotic Safety}\label{asym}

The scenario of asymptotically safe gravity, first proposed by Weinberg \cite{fixedpoint}, provides a fully renormalizable UV completion to gravity (for a review see \cite{Percacci:2007sz}). In this scenario, the dimensionless gravitational coupling approaches a non trivial fixed point in the UV.  The Planck mass is expected to become larger in the UV and the growth of amplitudes with energy of type $\xi^2 s/\bar M_P^2$ could be compensated by the running of the reduced Planck mass (see e.g. \cite{Hewett:2007st}). When gravity is coupled to matter the existence of the fixed point is even more difficult to establish, however detailed investigations have recently been carried out into scalar fields coupled to gravity \cite{Narain:2009fy, Narain:2009gb}. These studies incorporate the non-minimal coupling used in the original model of Higgs inflation \cite{Bezrukov:2007ep} and indicate that in the presence of these couplings a Gaussian matter fixed point could exist. A further result of their work is that if a non-trivial fixed point for gravity does exist, when scalar fields are introduced all the non-minimal couplings will be zero at the fixed point. This implies that $\xi$ gets smaller in the UV and would further counter the growth with energy of amplitudes.

Although asymptotically safe gravity can provide its own paradigm for inflation \cite{Weinberg:2009wa}, we remark here that should gravity display a fixed point in the presence of one of the matter couplings presented for Higgs inflation, Higgs inflation without any  physics beyond the standard model could be  free of unitarity problems. If we consider the running of coupling constants in the original example of Higgs inflation, where the cut off has been found to be $\bar M_P/\xi$, we would expect the Planck mass to increase and the non-minimal coupling to decrease at high energies, lifting the scale of unitarity violation as these energies are approached. Thus asympotically safe gravity with the standard model Higgs boson could provide a fully consistent inflationary scenario without having to introduce any new degrees of freedom. As suggested in \cite{Shaposhnikov:2009pv}, the Higgs boson's mass needs to be within a restricted mass range to avoid a Landau pole. If this is the case, the standard model with a large non-minimal coupling of the Higgs boson to the Ricci scalar can explain both the spontaneous symmetry breaking of the electroweak symmetry and inflation without the need for any new physics.

\section{Conclusions}

While the standard model Higgs boson would offer an elegant candidate for the inflaton there has been much debate about whether or not successful inflationary scenarios suffer from unitarity problems. Recent work suggested that the scale of unitarity violation was dependent on the background fields and since the Higgs field has a large background value during inflation, unitarity problems are in fact avoided throughout the inflationary period. Despite this, we point out that since new physics must appear well below the Planck mass in order to unitarize the model in today's universe, the original idea of the Higgs boson alone providing a paradigm for inflation has to be sacrificed. We would expect that either this new physics remains at the lower scale during inflation or any mechanism that lifts the mass scale would likely mix the new degrees of freedom with the Higgs boson. Either way, the new physics takes part in the inflationary dynamics.

A new model of Higgs inflation was proposed in \cite{Germani:2010gm} with the motivation being that it did not suffer from the unitarity problems that plagued the earlier model. Despite these claims, we find that when a careful analysis is carried out this model also suffers from unitarity problems. The scale of unitarity violation in this model is $(5 \bar H^2 M_P )^{1/3}$ and again this is lower than the size of the Higgs field during inflation, leading to the same unitarity problems and loss of control over the flat potential which troubled the earlier model.

We have also remarked that if gravity possesses a non trivial UV fixed point in the presence of a non-minimal coupling to the standard model Higgs boson, then the minimal Higgs inflation model could be self consistent without having to introduce new degrees of freedom.
\bigskip

\section*{Acknowledgments}

We would like to thank Mark Hindmarsh and Christoph Rahmede for helpful discussions.  The work of MA was supported by the Science and Technology Facilities Council [grant number ST/1506029/1].

%\newpage

%%%%%%%%%%%%%%%%%%%%%%%%%%%%%%%%%%%%%%%%%%%%%%%%%%%%%%%%%%%%%%%%%
%%%
%%%                     BIBLIOGRAPHY
%%%
%%%%%%%%%%%%%%%%%%%%%%%%%%%%%%%%%%%%%%%%%%%%%%%%%%%%%%%%%%%%%%%%%

\bigskip

%\newpage
%\vskip .75 in

\end{document}